\documentstyle[11pt]{article}

\newcounter{saveeqn}

\begin{document}

\title{Constraints on Space--Time Torsion \\ from Hughes--Drever  
Experiments}
\author{Claus L\"ammerzahl\thanks{e-mail:  
claus.laemmerzahl@uni-konstanz.de} \\ 
{\normalsize Fakult\"at f\"ur Physik, Universit\"at Konstanz,  
Postfach 5560 M674, D - 783434 Konstanz, Germany} \\ 
{\normalsize and} \\ 
{\normalsize Laboratoire de Gravitation et Cosmologie Relativiste,  
Universit\'e Pierre et Marie Curie,} \\ 
{\normalsize CNRS/URA 679, 75252 Paris Cedex 05, France}}

\maketitle

\begin{abstract}
The coupling of space--time torsion to the Dirac equation leads to  
effects on the energy levels of atoms which can be tested by  
Hughes--Drever type experiments. 
Reanalysis of these experiments carried out for testing the  
anisotropy of mass and anomalous spin couplings can lead to the till  
now tightest constraint on the axial torsion by $K \leq 1.5\cdot  
10^{-15}\;{\hbox{m}}^{-1}$. 
\end{abstract}


\section{Introduction}

The geometrical frame for General Relativity is a Riemannian  
space--time. 
Within this frame one can calculate solar system effects and finds  
within an accuracy of $10^{-4}$ that all predictions of GR are  
confirmed by experiment. 
The equivalence principle which is at the basis of the geometrisation  
of gravity is tested even to much better accuracy (for a review see  
\cite{Will93}).
However, on theoretical grounds this geometrical frame may be too  
narrow, and there are indeed many reasons to consider a more general  
geometrical structure as mathematical description of physical  
space--time. 
One very prominent generalisation is the Riemann--Cartan geometry  
which 
(i) is the most natural generalisation of a Riemannian geometry by  
allowing a non--symmetric metric--compatible connection, (ii) treats  
spin on the same level as mass as it is indicated by the group  
theoretical analysis of the Poincar\'e group, and (iii) arises in most  
gauge theoretical approaches to General Relativity, as e.g. in the  
Poincar\'e--gauge theory  
\cite{HehlHeydeKerlick76,HehlLemkeMielke91,AHL92} or supergravity  
\cite{Nieuwenhuizen81}. 
However, till now there is no experimental evidence for torsion. 
On the other hand, from the lack of effects which may be due to  
torsion one can calculate estimates on the maximal strength of the  
torsion fields. 
This is the purpose of this Letter: We first calculate that torsion  
in principle influences the experimental outcome of Hughes--Drever  
type experiments (for a review on these experiments see  
\cite{Will93}). 
Since no effects were observed we get from the accuracy of these  
experiments upper bounds on the torsion strength. 
Therefore, by means of a reinterpretation of the Hughes--Drever type  
experiments we obtain the till now most stringent upper bounds on the  
torsion strength.

While torsion does not influence the behaviour of macroscopic bodies  
\cite{StoegerYasskin79,YasskinStoeger80} it acts on the evolution of  
spin degrees of freedom and can in principle be measured by  
determining the precession of an elementary spin \cite{Audretsch83}. 
This spin--torsion interaction also modifies in first order of  
$\hbar$ the trajectory of an elementary particle. 
Turning around the way of reasoning, it is also possible to establish  
torsion by allowing the spin to behave in a way not predicted by  
General Relativity \cite{AL88,AL93}. 
The effect of torsion on spin can also influence the outcome of an  
interference experiment with neutrons \cite{AL83,AnandanLesche83}. 
This fact and the lack of any experimental evidence for a coupling  
has been used in Ref.\ \cite{AnandanLesche83} to pose an upper bound  
of $\leq 10^{-7}\;{\hbox{m}}^{-1}$ on the strength of torsion fields. 
Another estimate \cite{deSabbataetal91} relates the strength of the  
coupling of spinning matter to torsion to the density of polarised  
particles. 
We will show that by means of a reinterpretation of Hughes--Drever  
type experiments we can restrict torsion to $\leq  
10^{-15}\;{\hbox{m}}^{-1}$. 
This of course does not mean that torsion does not exist or does not  
play an important role in our understanding of gravitation.

In the following we first derive the non--relativistic limit of the  
Dirac equation in a Riemann--Cartan space--time. 
The resulting Pauli--equation with additional coupling to the axial  
torsion vector is used to derive the Hamiltonian for the energy  
levels of a bound two--particle system.
This determines in the non--relativistic regime the energy levels of  
a nucleus consisting in a core and one valence proton which is the  
physical system usually taken to perform Hughes--Drever type  
experiments.

A Riemann--Cartan space--time consists in a metric $g_{\mu\nu}$  
($\mu, \nu = 0, \ldots, 3$) and a metric--compatible connection  
$\Gamma_{\mu\nu}^\sigma = \{{\sigma\atop{\mu\nu}}\} -  
K_{\mu\nu}^{\phantom{\mu\nu}\sigma}$ where  
$\{{\sigma\atop{\mu\nu}}\}$ is the usual Christoffel connection and  
$K_{\mu\nu}^{\phantom{\mu\nu}\sigma}$ is the contorsion tensor  
related to the torsion $S_{\mu\nu}^{\phantom{\mu\nu}\sigma}$ through  
$S_{\mu\nu}^{\phantom{\mu\nu}\sigma} = \Gamma_{[\mu\nu]}^\sigma = -  
K_{\mu\nu}^{\phantom{\mu\nu}\sigma}$. 
We introduce tetrads $h_a^\mu$ ($a, b = 0, \ldots, 3$) through  
$g_{\mu\nu} h_a^\mu h_b^\nu = \eta_{ab}$ with $\eta = \hbox{diag}(- +  
+ +)$.

\section{Dirac equation in Riemann-Cartan space--time}

The Dirac matrices $\gamma^\mu$ are defined by the Clifford algebra  
$\gamma^{(\mu} \gamma^{\nu)} = g^{\mu\nu}$ and are connected with the  
standard Minkowski Dirac matrices $\gamma^a$ fulfilling $\gamma^{(a}  
\gamma^{b)} =\eta^{ab}$ by $\gamma^\mu = h^\mu_a \gamma^a$. 
In Dirac representation ($m = 1, 2, 3$)
\begin{equation}
\gamma^{(0)} = - i \beta, \; \gamma^m = - i \beta \alpha^m
\end{equation}
with 
\begin{eqnarray}
\alpha^m & = & \left(\matrix{0 & \sigma^m \cr \sigma^m &  
0\cr}\right), \qquad
\beta = \left(\matrix{1 & 0\cr 0 & - 1\cr}\right) \\
\Sigma^m & = & \left(\matrix{\sigma^m & 0\cr 0 & \sigma^m\cr}\right),  
\qquad 
\gamma_5 = \left(\matrix{0 & 1\cr 1 & 0\cr}\right)\, .
\end{eqnarray}
We use the Dirac equation which is derived from a minimally coupled  
Dirac Lagrangian \cite{HehlDatta71,HehlHeydeKerlick76,Audretsch83} 
(for a recent review see also \cite{Mielkeetal96})
\begin{eqnarray}
0 & = & i \hbar \gamma^\mu D_\mu\psi + {i\over 2} 
K_{\rho\mu}^{\phantom{\rho\mu}\rho} \gamma^\mu \psi + m c \psi 
\nonumber \\ 
& = & i \hbar \gamma^\mu {\buildrel {\{\}} \over D}_\mu\psi - \hbar  
K_\mu \gamma_5 \gamma^\mu \psi + m c \psi\; ,
\end{eqnarray}
where
\begin{equation}
D_\mu \psi = \partial_\mu \psi + \Gamma_\mu \psi 
\end{equation}
with the spinorial representation of the anholonomic connection
\begin{eqnarray}
\Gamma_\mu & = & {1\over 4} D_\mu h_a^\nu h^b_\nu \gamma_b \gamma^a  
\nonumber\\
& = & {1\over 4} {\buildrel {\{\}} \over D}_\mu h_a^\nu h^b_\nu  
\gamma_b \gamma^a - {1\over 4} K_{\mu\sigma}^{\phantom{\mu\sigma}\nu}  
h_a^\sigma h^b_\nu \gamma_b \gamma^a \, .
\end{eqnarray}
${\buildrel {\{\}} \over D}_\mu$ and ${\buildrel {\{\}} \over  
\Gamma}_\mu$ is the Christoffel part of the covariant derivative and  
connection, respectively. 
$K_\mu = {1\over 6} \epsilon_{\mu\sigma}^{\phantom{\mu\sigma}\nu\rho}  
K_{\nu\rho}^{\phantom{\nu\rho}\sigma}$ is the axial part of the  
space--time torsion.

\section{Non--relativistic limit: Pauli equation in Riemann-Cartan  
space--time}

In order to carry through a non--relativistic approximation of the  
Dirac equation coupled to metric and torsion we first perform a  
Newtonian approximation of Riemann--Cartan theory. 
Along the lines of \cite{FischbachFreemanCheng81} we expand the  
metric with respect to the Newtonian potential in a quasi--Newtonian  
coordinate system ($dx^0 = c\, dt$, $i, j = 1, 2, 3$) 
\begin{eqnarray}
g_{00} & = & - 1 + 2 {U\over{c^2}} \label{metric00} \\ 
g_{0i} & = & 0 \label{metric0i} \\ 
g_{ij} & = & \left(1 + 2 {U\over{c^2}}\right) \delta_{ij} \, .
\end{eqnarray}
The corresponding tetrads are
\begin{eqnarray}
h_{(0)}^0 & = & 1 + {U\over{c^2}} \\ 
h_{(0)}^i & = & 0 \\ 
h_m^0 & = & 0 \\ 
h_m^i & = & \left(1 - {U\over{c^2}}\right) \delta_m^i 
\end{eqnarray}
from which one can calculate the matrices $\gamma^\mu$.
The Riemannian part of the spinorial representation of the anholonomic 
connection is
\begin{eqnarray}
{\buildrel {\{\}} \over \Gamma}_0 & = & - {1\over{2 c^2}} \alpha^i  
\partial_i U \\ 
{\buildrel {\{\}} \over \Gamma}_i & = & - {i\over{2 c^2}}  
\epsilon_{ij}^{\phantom{ij}k} \Sigma^j \partial_k U \, .
\end{eqnarray}
The additional axial torsion is taken to be approximately constant at  
the position of the experiment.

We insert the metric, tetrads and connection into the Dirac equation  
and solve it with respect to ${\partial\over{\partial t}}\psi$ where  
we neglect squares of the Newtonian potential and products of $U$ and  
torsion
\begin{equation}
i \hbar {\partial\over{\partial t}} \psi = - i \hbar c \left(1 - 2  
{U\over{c^2}}\right) \alpha^i \partial_i \psi -  {{i \hbar}\over{2  
c}} \alpha^i \partial_i U \psi + \hbar c K_{(0)} \gamma_5 \psi -  
\hbar c K_i \Sigma^i \psi + \left(1 - {U\over{c^2}}\right) \beta m  
c^2 \psi
\end{equation}
In order to perform a Foldy--Wouthuysen transformation of the  
Hamiltonian  (compare \cite{BjorkenDrell64}) we split the Hamiltonian  
into even and odd parts:
\begin{equation}
H = {\cal O} + {\cal E} + \beta m c
\end{equation}
with
\begin{eqnarray}
{\cal O} & = & - i \hbar c \left(1 - 2 {U\over{c^2}}\right) \alpha^i  
\partial_i -  {{i \hbar}\over{2 c}} \alpha^i \partial_i U + \hbar c  
K_{(0)} \gamma_5 \\ 
{\cal E} & = & \left(1 - {U\over{c^2}}\right) \beta m c^2 - \hbar c  
K_i \Sigma^i 
\end{eqnarray}
and get
\begin{eqnarray}
H^\prime\varphi & = & \beta \left(m c^2 + {{{\cal O}^2}\over{2 m  
c^2}} - {{{\cal O}^4}\over{8 m^3 c^6}}\right)\varphi + {\cal  
E}\varphi - {1\over{8 m^2 c^4}} \left[{\cal O}, [{\cal O}, {\cal  
E}]\right]\varphi - {i\over{8 m^2 c^4}} [{\cal O}, \dot{\cal  
O}]\varphi \nonumber\\
& = & \beta \left(m c^2 \varphi + \left(- {{\hbar^2}\over{2 m}}\Delta  
\varphi - {{\hbar}\over{m}} K_{(0)} \Sigma^i i \hbar \partial_i  
\varphi\right)\right) - U \beta m \varphi - \hbar c K_i \Sigma^i
\end{eqnarray}
where we neglected all relativistic corrections and squares and  
derivatives of the torsion. 
Projection onto the large components and elimination of the rest  
energy gives
the Pauli--equation in a Riemann--Cartan space--time 
\begin{equation}
i \hbar {\partial\over{\partial t}} \psi = - {{\hbar^2}\over{2  
m}}\Delta \psi - {{\hbar}\over{m}} K_{(0)} \sigma^i i \hbar  
\partial_i \psi - m U \psi - \hbar c K_i \sigma^i \psi \, .  
\label{PauliEqu}
\end{equation}
(The occurence of $c$ in the last term just means that $K_i$ has the  
dimension 1/length.)  
The coupling to the electromagnetic field is accomplished by 
\begin{equation}
i \hbar {\partial\over{\partial t}} \psi = - {{\hbar^2}\over{2  
m}}\nabla^2 \psi - e \phi \psi - {{e \hbar}\over{2 m c}} H_i \sigma^i  
\psi - m U \psi - {{\hbar}\over{m}} K_{(0)} \sigma^i i \hbar \nabla_i  
\psi - \hbar c K_i \sigma^i \psi = H \psi
\end{equation}
with $\nabla_i = \partial_i - {{ie}\over{\hbar c}} A_i$. 
The coupling to rotation $\Omega_i$ can be introduced either by  
starting with the appropriately modified metric components  
(\ref{metric00}) and (\ref{metric0i}) (see \cite{NiZimmermann78}) or  
by performing a unitary transformation of the quantum states obeying  
(\ref{PauliEqu}) to a rotating frame \cite{Mashhoon88}, see also  
\cite{Mashhoon95} and references therein. 
In each case one gets the additional coupling term $\Omega_i (L^i +  
{1\over 2} \hbar \sigma^i)$ where $L^i = \epsilon^{ijk} x^j (- i  
\hbar \nabla_k)$ is the angular momentum operator. 
(Since we are treating the non--relativistic limit without second quantisation, 
we encounter no problems which may arise from a non--renormalisability of 
a Fermi type coupling when inserting into the 
Dirac equation the field equations for torsion 
within the Einstein--Cartan theory \cite{HehlHeydeKerlick76}.)

\section{The two--particle--Hamiltonian}

The two--particle Hamiltonian is given by the sum of two Hamiltonians  
of the above form. 
We restrict from the very beginning to the following case given by  
the experiment: 
A nucleus consisting in a core with vanishing total angular momentum  
($J = 0$) and a valence proton with spin $S = {1\over 2}$ and some  
angular momentum $L$. 
The wave function is a function of two position variables and an  
angular momentum variable $\psi = \psi_{J, M_J}(x_1, x_2, t)$. 
Then 
\begin{equation}
H = H_1 + H_2 + V \label{Hamtot}
\end{equation}
with
\begin{eqnarray}
H_1 & = & - {{\hbar^2}\over{2 m_1}}\nabla^2_1 - e_1 \phi - m_1 U \\ 
H_2 & = & - {{\hbar^2}\over{2 m_2}}\nabla^2_2 - e_2 \phi- m_2 U -  
\mu_2 H_i \sigma^i - {{\hbar}\over{m_2}} K_{(0)} \sigma^i i \hbar  
\nabla_{2i} - \hbar c K_i \sigma^i
\end{eqnarray}
where $V = V(x_2 - x_1)$ is some binding potential, $\mu_2 = e_2  
\hbar/2 m_2 c$, and $\nabla_{1m}$ denotes the $U(1)$--covariant  
derivative with respect to the coordinate $x_1^m$. 
The quantisation axis is defined by the external magnetic field  
$H_i$.

We introduce the relative coordinate $x$ and a center-of-mass  
coordinate $X$
\begin{equation}
x := x_2 - x_1, \qquad X := {{m_1}\over{m_1 + m_2}} x_1 +  
{{m_2}\over{m_1 + m_2}} x_2 
\end{equation} 
and insert the corresponding coordinate transformation into the  
Hamiltonian (\ref{Hamtot}). 
We also split the electromagnetic potentials $A_i$ and $\phi$ into a  
part due to external sources $A_i^{\hbox{\scriptsize e}}$,  
$\phi^{\hbox{\scriptsize e}}$ and a part due to the charge of the  
other particle $A_i^{\hbox{\scriptsize i}}$, $\phi^{\hbox{\scriptsize  
i}}$: $A_i = A_i^{\hbox{\scriptsize e}} + A_i^{\hbox{\scriptsize  
i}}$, $\phi = \phi^{\hbox{\scriptsize e}} + \phi^{\hbox{\scriptsize  
i}}$. 
We absorb the electromagnetic field which is created by the particles  
itself and which is not connected with a derivative into a modified  
potential $V^\prime(x)$ which depends on the relative coordinate  
only. 
Then the transformed two--particle Pauli equation reads (in the  
following we omit the index $J, M_J$ characterising the wave  
function)
\begin{eqnarray}
H \varphi & = & - {{\hbar^2}\over{2 m}} \Delta_X \varphi -  
{{\hbar^2}\over{2 m_{\hbox{\scriptsize red}}}} \Delta_x \varphi + {{i  
\hbar}\over{c}} \left({{e_2}\over{m_2}} A^{\hbox{\scriptsize  
e}}_i(x_2) - {{e_1}\over{m_1}} A^{\hbox{\scriptsize e}}_i(x_1)\right)  
{{\partial\varphi}\over{\partial x^i}} \nonumber\\
& &  + {{i \hbar}\over{m c}} \left(e_1 A^{\hbox{\scriptsize  
e}}_i(x_1) + e_2 A^{\hbox{\scriptsize e}}_i(x_2)\right) {{\partial  
\varphi}\over{\partial X^i}} + {{i \hbar}\over{c}}  
A^{\hbox{\scriptsize i}}_j(x) {e\over{m}} {{\partial  
\varphi}\over{\partial X^j}}  \nonumber\\
& & + {{i \hbar}\over{c}} A^{\hbox{\scriptsize i}}_j(x)  
\left({{e_2}\over{m_2}} - {{e_1}\over{m_1}}\right)  
{{\partial\varphi}\over{\partial x^j}} - \mu_2 H_i \sigma^i \varphi +  
m_2 U(x_2)\varphi + m_1 U(x_1) \varphi \nonumber\\
& & - \hbar c K_i \sigma^i \varphi - i {{\hbar^2}\over{m_2}} K_{(0)}  
\sigma^i \left({{m_2}\over m} {\partial\over{\partial X^i}} +  
{\partial\over{\partial x^i}}\right) \varphi  \nonumber\\
& & + V^\prime(x) \varphi + e_1 \phi^{\hbox{\scriptsize e}}(x_1)  
\varphi + e_2 \phi^{\hbox{\scriptsize e}}(x_2) \varphi 
\end{eqnarray}
where we defined the total charge $e := e_1 + e_2$, the total mass $m  
:= m_1 + m_2$, the reduced mass $m_{\hbox{\scriptsize red}} = m_1  
m_2/m$, used the Coulomb gauge for the external electromagnetic  
potential, and neglected squares of the electromagntic potential and  
products of torsion with the electromagnetic potential. 
We approximate 
\begin{eqnarray}
{{e_2}\over{m_2}} A^{\hbox{\scriptsize e}}_i(x_2) - {{e_1}\over{m_1}}  
A^{\hbox{\scriptsize e}}_i(x_1) & \approx & {{e_{\hbox{\scriptsize  
red}}}\over{m_{\hbox{\scriptsize red}}}} A^{\hbox{\scriptsize  
e}}_i(X) + {{2 c}\over\hbar} \mu_{\hbox{\scriptsize red}} x^k  
\nabla_k A^{\hbox{\scriptsize e}}_i(X) \\
e_1 A^{\hbox{\scriptsize e}}_i(x_1) + e_2 A^{\hbox{\scriptsize  
e}}_i(x_2) & \approx & e A^{\hbox{\scriptsize e}}_i(X) +  
e_{\hbox{\scriptsize red}} x^k \nabla_l A^{\hbox{\scriptsize e}}_i(X)  
\\
e_1 \phi(x_1) + e_2 \phi(x_2) & \approx & e \phi(X) +  
e_{\hbox{\scriptsize red}} x^i \nabla_i \phi(X) \\
m_1 U(x_1) + m_2 U(x_2) & \approx & m U(X) + {1\over 2}  
m_{\hbox{\scriptsize red}} x^k x^l \nabla_k \nabla_lU(X)
\end{eqnarray}
where we introduced the `reduced charge
\begin{equation}
e_{\hbox{\scriptsize red}} := m_{\hbox{\scriptsize red}}  
\left({{e_2}\over{m_2}} - {{e_1}\over{m_1}}\right)
\end{equation} 
and the `reduced Bohr's magneton'
\begin{equation}
\mu_{\hbox{\scriptsize red}} := {\hbar\over{2 m c}} \left(m_1  
{{e_2}\over{m_2}} + m_2 {{e_1}\over{m_1}}\right)\, .
\end{equation}
(In the usually considered case $m_1 \rightarrow \infty$ (very heavy  
nucleus) we get $e_{\hbox{\scriptsize red}} \rightarrow e_2$ and  
$\mu_{\hbox{\scriptsize red}} \rightarrow e_2 \hbar/2 m_2 c  
= \mu_2$.)
If we complete the partial derivatives to covariant derivatives,  
neglect squares of the Maxwell potential, and neglect the internal  
vector potential $A^{\hbox{\scriptsize i}}_j(x)$ which is of the  
order $c^{-1}$, then we get
\begin{eqnarray}
H \varphi & = & - {{\hbar^2}\over{2 m}} \nabla_X^2 \varphi -  
{{\hbar^2}\over{2 m_{\hbox{\scriptsize red}}}}  \Delta_x \varphi +  
{{e_{\hbox{\scriptsize red}}}\over{m c}} x^l \nabla_l  
A^{\hbox{\scriptsize e}}_j(X) i \hbar {{\partial  
\varphi}\over{\partial X^j}} \nonumber\\
& & + 2 \mu_{\hbox{\scriptsize red}} x^k \nabla_k  
A^{\hbox{\scriptsize e}}_j(X) i {{\partial\varphi}\over{\partial  
x^j}} - \mu_2 H_i \sigma^i \varphi + V^\prime(x) \varphi \nonumber\\
& & +  e \phi^{\hbox{\scriptsize e}}(X) \varphi +  
e_{\hbox{\scriptsize red}} x^i \nabla_i \phi^{\hbox{\scriptsize  
e}}(X) \varphi + m U(X) \varphi + {1\over 2} m_{\hbox{\scriptsize  
red}} x^k x^l \nabla_k \nabla_l U(X) \varphi  \nonumber\\
& & - i {{\hbar^2}\over{m_2}} K_{(0)} \sigma^i \left({{m_2}\over m}  
{\partial\over{\partial X^i}} + {\partial\over{\partial x^i}}\right)  
\varphi - \hbar c K_i \sigma^i \varphi \label{Ham}
\end{eqnarray}
with $\nabla_{Xi} = {\partial\over{\partial X^i}} - {{i e}\over{\hbar  
c}} A^{\hbox{\scriptsize e}}_i(X)$. 
This is the final form of the Hamilton operator expressed with  
respect to the relative coordinates $x$ and the center--of--mass  
coordinates $X$.

Next we extract from this total Hamiltonian that Hamiltonian which  
describes the energy levels of this bound system by freezing the  
center--of--mass motion and keeping the center--of--mass coordinate  
of the atom at $X = 0$. 
In addition, we specialise to $\phi^{\hbox{\scriptsize e}} = 0$ and  
gauge away constant terms so that we get
\begin{eqnarray}
H \varphi & = & - {{\hbar^2}\over{2 m_{\hbox{\scriptsize red}}}}   
\Delta_x \varphi + 2 \mu_{\hbox{\scriptsize red}} x^k \nabla_k  
A^{\hbox{\scriptsize e}}_i(0) i \hbar  
{{\partial\varphi}\over{\partial x^i}} - \mu_2 H_i \sigma^i \varphi   
+ V^\prime(x) \varphi \nonumber\\
& & + {1\over 2} m_{\hbox{\scriptsize red}} x^k x^l \nabla_k \nabla_l  
U(0) \varphi - i {{\hbar^2}\over{m_2}} K_{(0)} \sigma^i  
{\partial\over{\partial x^i}} \varphi - \hbar c K_i \sigma^i \varphi  
\, .
\end{eqnarray}
Again we neglected squares of the vector potential. 
We take a constant external magnetic field: $A_i^{\hbox{\scriptsize  
e}}(X) = {1\over 2} \epsilon_{ilk} H^l X^k$. 
Then the  Hamiltonian giving the energy levels is
\begin{eqnarray}
H_{\hbox{\scriptsize E}} & = & - {{\hbar^2}\over{2  
m_{\hbox{\scriptsize red}}}} \Delta_x - \mu_{\hbox{\scriptsize red}}  
H^l \delta^{ij} \epsilon_{ilk} x^k i \hbar {{\partial}\over{\partial  
x^j}} - \mu_2 H_i \sigma^i  \nonumber\\
& & - {{\hbar^2}\over{m_2}} K_{(0)} \sigma^i i  
{{\partial}\over{\partial x^i}} + {1\over 2} m_{\hbox{\scriptsize  
red}} x^k x^l \nabla_k \nabla_lU(0) - \hbar c K_i \sigma^i +  
V^\prime(x) \, .\label{finalHam}
\end{eqnarray}
If we consider rotation we have to add $\Omega_i (l^i + {1\over 2}  
\hbar \sigma^i)$ where $l^i$ is the angular momentum with respect to  
the relative coordinates. 
The spin--rotation term has been dicussed by Mashhoon  
\cite{Mashhoon88,Mashhoon95} and the $\Omega_i l^i$ term by Silverman  
\cite{Silverman89}. 
We get various parts for this Hamiltonian describing the energy  
levels of a bound system: 
\begin{equation}
H_{\hbox{\scriptsize E}} = H_0 + H_{\hbox{\scriptsize em}} +  
H_{\hbox{\scriptsize Newton}} + H_{\hbox{\scriptsize torsion}}
\end{equation}
with
\begin{eqnarray}
H_0 & = & - {1\over{2 m_{\hbox{\scriptsize red}}}} \Delta +  
V^\prime(x) \\
H_{\hbox{\scriptsize em}} & = & - H_i \left({{\mu_{\hbox{\scriptsize  
red}}}\over\hbar} l^i + \mu_2 \sigma^i\right) \\
H_{\hbox{\scriptsize Newton}} & = & {1\over 2} m_{\hbox{\scriptsize  
red}} x^k x^l \nabla_k \nabla_lU(0)  \\
H_{\hbox{\scriptsize torsion}} & = & - {{\hbar^2}\over{m_2}} K_{(0)}  
\sigma^i i {{\partial}\over{\partial x^i}} - \hbar c K_i \sigma^i  
\label{HnonEinst}
\end{eqnarray}
For the electric proton--nucleus interaction we have $V^\prime(x) = -  
Z e^2/x$. 
For the nuclear proton--nucleus interaction we have to take some  
appropriate model for the potential of the nucleus, e.g. the harmonic  
oscillator potential or Wood--Saxon potential. 
Note that there are no Einsteinian effects due to the acceleration  
$\nabla U$. 
This is in agreement with the equivalence principle: The effect of  
gravitational acceleration can be cancelled by a transformation to a  
suitable accelerated frame and therefore does not influence the  
energy levels. 
$H_0$ describes the atom without external fields,  
$H_{\hbox{\scriptsize em}}$ the Zeeman effect. 
The third Hamiltonian is the usual gravitational interaction with the  
Newtonian part of the Riemannian space--time curvature. 
The last term describes the coupling to torsion under consideration. 
The first term amounts to a spin--momentum coupling which will lead  
to second order effects only. 
This is the generalised Pauli--equation for the energy levels in a  
Riemann--Cartan space--time.

\section{Comparison with experiment}

We use the above Hamiltonian to calculate the Zeeman--splitting of  
energy levels in an atom  in the presence of torsion.  
We describe the case which is considered in usual Hughes-Drever type  
experiments (see e.g.  \cite{Will93}) namely an atomic nucleus which  
consists in a $J = 0$ core and a valence proton with angular momentum  
$L = 1$. 
Our quantisation axis for the spin is given by the external magnetic  
field $H_i$ (we can now use $\mu_{\hbox{\scriptsize red}} \approx  
\mu_2 = \mu_{\hbox{\scriptsize B}}$). 
We are going to calculate the shifts in the energy levels due to the  
interaction Hamiltonian (\ref{HnonEinst}) describing non--Einsteinian  
effects. 
We use first order perturbation theory. 
The unperturbed states $|J, M_J\rangle$ are given by 
\begin{equation}
|{\textstyle {3\over 2}, {3\over 2}}\rangle = \left(\matrix{|1,  
1\rangle  \cr 0 \cr}\right), \quad
|{\textstyle {3\over 2}, {1\over 2}}\rangle =  
\left(\matrix{\sqrt{2\over 3} |1, 0\rangle \cr \sqrt{1\over 3} |1,  
1\rangle  \cr}\right), \quad
|{\textstyle {3\over 2}, - {1\over 2}}\rangle =  
\left(\matrix{\sqrt{1\over 3} |1, -1\rangle \cr \sqrt{2\over 3} |1,  
0\rangle \cr}\right), \quad
|{\textstyle {3\over 2}, - {3\over 2}}\rangle = \left(\matrix{0 \cr  
|1, - 1\rangle\cr}\right)
\end{equation}
The interaction Hamiltonian under consideration  (\ref{HnonEinst})  
has the structure $({\cal A}_k^i p_i + {\cal A}_k) \sigma^k$. 
We neglect effects due to $\nabla_i\nabla_j U$ since these effects  
give energy shifts smaller than $10^{-40}\;\hbox{eV}$ which is too  
small to be detectable.
We get for the corresponding expectation values
\begin{eqnarray}
\langle {\textstyle {3\over 2}, {3\over 2}}| {\cal A}_k^i p_i + {\cal  
A}_k |{\textstyle {3\over 2}, {3\over 2}}\rangle & = & \langle 1, 1|  
{\cal A}_z |1, 1\rangle  \\
\langle {\textstyle {3\over 2}, {1\over 2}}| {\cal A}_k^i p_i + {\cal  
A}_k |{\textstyle {3\over 2}, {1\over 2}}\rangle & = & {2\over 3}  
\langle 1, 0| {\cal A}_z |1, 0\rangle - {1\over 3} \langle 1, 1|  
{\cal A}_z |1, 1\rangle \\
\langle {\textstyle {3\over 2}, - {1\over 2}}| {\cal A}_k^i p_i +  
{\cal A}_k |{\textstyle {3\over 2}, - {1\over 2}}\rangle & = &  
{1\over 3} \langle 1, -1| {\cal A}_z |1, -1\rangle - {2\over 3}  
\langle 1, 0| {\cal A}_z |1, 0\rangle \\
\langle {\textstyle {3\over 2}, -{3\over 2}}| {\cal A}_k^i p_i +  
{\cal A}_k |{\textstyle {3\over 2}, -{3\over 2}}\rangle & = & -  
\langle 1, -1| {\cal A}_z |1, -1\rangle 
\end{eqnarray}
where we used that the expectation value for expressions linear in  
the momentum vanishes. The transition frequencies turn out to be 
\begin{eqnarray}
\hbar \omega({\textstyle {3\over 2} \rightarrow {1\over 2}}) & = &  
{4\over 3} \langle 1, 1| {\cal A}_z  |1, 1\rangle - {2\over 3}  
\langle 1, 0| {\cal A}_z  |1, 0\rangle  \\
\hbar \omega({\textstyle {1\over 2} \rightarrow - {1\over 2}}) & = &  
{4\over 3} \langle 1,0| {\cal A}_z  |1,0\rangle - {1\over 3} \langle  
1, 1| {\cal A}_z  |1, 1\rangle - {1\over 3} \langle 1, -1| {\cal A}_z   
|1, -1\rangle  \\
\hbar \omega({\textstyle -{1\over 2} \rightarrow -{3\over 2}}) & = &  
{4\over 3} \langle 1, -1| {\cal A}_z  |1, -1\rangle - {2\over 3}  
\langle 1, 0| {\cal A}_z  |1, 0\rangle \, .
\end{eqnarray}
The matrix elements are $\langle 1, 1| {\cal A}_k |1, 1\rangle =  
\langle 1, 0| {\cal A}_k |1, 0\rangle = \langle 1, -1| {\cal A}_k |1,  
-1\rangle = {\cal A}_z$ so that we get an equal shift $\hbar\omega =  
{2\over 3} \hbar c K_z$ for all three transition frequencies.
The search for such a shift during the change of the $z$-axis with  
respect to the orthogonal nonrotating Newtonian coordiante system  
amounts to a Hughes-Drever experiment. 
If space--time torsion will be detected it will lead to a diurnal  
shift of the Zeeman singlet line. 
However, present experiments (see  
\cite{Chuppetal89,Prestageetal85,Lamoreauxetal86}) didn't detect any  
effects. 
Indeed, the experimental setup of Chupp {\it et al}.\  
\cite{Chuppetal89} uses two kinds of atoms, ${}^{21}{\hbox{Ne}}$ and  
${}^3{\hbox{He}}$, where the latter serves as magnetometer standard. 
Both kinds of atoms are subject to the same magnetic field which can  
be controlled to an accuracy of $\delta B \leq 10^{-10}\;\hbox{G}$. 
Since both atoms possess different $g$--factors,  
$g({}^{21}{\hbox{Ne}}) = - 0.6619 \mu_{\hbox{\scriptsize B}}$ and  
$g({}^3{\hbox{He}}) = - 2.1276 \mu_{\hbox{\scriptsize B}}$ where  
$\mu_{\hbox{\scriptsize B}}$ is Bohr's magneton of a nucleon, the  
Zeeman lines are different. 
During the experiment the energy difference of these two Zeeman  
frequencies 
\begin{equation}
E({}^3{\hbox{He}}) - E({}^{21}{\hbox{Ne}}) = \left(g({}^3{\hbox{He}})  
- g({}^{21}{\hbox{Ne}})\right) (\mu_{\hbox{\scriptsize B}} B - \hbar  
c K_z)
\end{equation}
can be recorded. 
Here $K_z$ is a function of the orientation of the quantisation axis. 
If this axis is fixed to the surface of the earth it can exhibit a  
diurnal time dependence. 
The accuracy of the experiments can be described in terms of an  
effective variation in the magnetic field $B$, $\delta B \leq  
10^{-10}\:\hbox{G}$. 
Consequently, if the influence of torsion $\hbar c K_z$ is larger  
than $\mu_{\hbox{\scriptsize B}} \delta B$, then an observable effect  
would occur.
If one is going to redo this experiment, searches for the above  
described effect and observes a null--result, then we are lead to the  
following estimate on torsion
\begin{equation}
K_z \leq {1\over{\hbar c}} \mu_{\hbox{\scriptsize B}} \delta B \leq  
1.5\cdot 10^{-15}\;{\hbox{m}}^{-1} \, .
\end{equation}
From this new version of an already performed Hughes--Drever like  
experiment we may get the up to now best estimate for the space--components  
of the axial part of a hypothetical space--time torsion. 
It is not possible to test other parts of the torsion tensor since  
the Dirac equation couples to the axial part only. 
One needs higher spin equations for a coupling to the trace and the  
traceless part of the torsion tensor.

Originally Hughes--Drever type experiments are designed to search for  
possible anisotropies of space--time, or, equivalently, anisotropies  
of the mass of quantum systems. 
These parts will lead to a splitting of the singlet line to a triplet  
line. 
Therefore, it is possible to distinguish between effects due to mass  
anisotropy and torsion: while the first cause leads to a splitting of  
the singlet line, the torsion shifts the whole line spectrum in the same  
way.

Also the experiments designed to search for an anomalous  
spin--coupling \cite{Venemaetal92} can be used for estimating the  
strength of the torsion coupling. 
This experiment has been analysed by Mashhoon \cite{Mashhoon95} in  
order to show that it tests indirectly the spin--rotation coupling. 
Since space--time torsion couples to the spin in the same way as  
rotation (\ref{finalHam}), one also can draw the same conclusions  
regarding torsion. 
One arrives at similar estimates as above.

Although $K_{(0)}$ leads to second order effects only in the energy  
shift, it influences the center--of--mass motion via a spin--momentum  
coupling (compare eqn (\ref{Ham})). 
Such a coupling can be tested with atom beam interferometry. 
Using  a spin flip as described in \cite{Mashhoon88} or in  
\cite{ABL93} for testing other spin--momentum couplings, we get the  
phase shift
\begin{equation}
\delta\phi = K_{(0)} \Delta l 
\end{equation}
where $\Delta l$ is the distance between splitting and recombination  
of the atomic beam. 
Note that this phase shift is nondispersive and does not depend on  
the interaction time. 
If we take an absolute accuracy $\delta\phi \leq 10^{- 2}$ and  
$\Delta l = 1 \,\hbox{m}$ and assume a null experiment, then we get  
the estimate $K_{(0)} \leq 10^{-2} {\hbox{m}}^{-1}$ for the time  
component of the axial torsion. 
The above phase shift which comes from a spin--momentum coupling due  
to the existence of torsion can be distinguished through its mass  
independence from a similar phase shift due to a violation of local  
Lorentz invariance \cite{ABL93}.

\section*{Acknowledgement}

I want to thank Prof.\ Ch.J.\ Bord\'e, Prof.\ F.W. Hehl and Dr.\ St.\  
Schiller for disussions, Prof.\ R.\ Kerner for the hospitality at the  
LGCR, Paris, as well as the Deutsche Forschungsgemeinschaft and the  
CNRS (France) for financial support.

\end{document}